# Observation of re-entrant spin glass behavior in $(Ce_{1-x}Er_x)Fe_2$ compounds

Arabinda Haldar[1], K. G. Suresh[1,*] and A. K. Nigam[2]

[1]Magnetic Materials Laboratory, Department of Physics, Indian Institute of Technology Bombay, Mumbai- 400076, India

[2]Tata Institute of Fundamental Research, Homi Bhabha Road, Mumbai- 400005, India

*Abstract*

Clear experimental evidence of re-entrant spin glass state has been revealed in Er doped $CeFe_2$ compounds. The zero field cooled - field cooled bifurcation in dc magnetization, frequency dependence of freezing temperature, relaxation in zero field cooled magnetization and presence of large remanence confirm the spin glass state in these compounds. Frequency dependence is found to follow the critical slowing down mechanism of the type: $\tau = \tau_0 (T_f / T_{SG} - 1)^{-zv}$. The random substitution of Er and the change in the valence state of Ce along with an enhancement of the ferromagnetic component in the Fe sublattice seem to be responsible for the spin glass state. Using detailed experimental protocols, we also prove that the low temperature state in these compounds is not a magnetic glass. The absence of exchange bias gives an indication that there is no coexistence of ferromagnetism and spin glass state in these compounds. The RSG state is found to be associated with the randomly magnetized clusters instead of atomic level randomness.

*Corresponding author (email: suresh@phy.iitb.ac.in)

## I. Introduction

Among the $R$Fe$_2$ ($R$ = rare-earth element) series of compounds, CeFe$_2$ has attracted a lot of attention from various researchers in view of its anomalous magnetic properties such as low Curie temperature ($T_C$ = 230 K), low saturation magnetic moment ($M_S$ = 2.4 $\mu_B / f.u.$) as compared to that of LuFe$_2$ ($T_C$ = 610 K, $M_S$ = 2.9 $\mu_B / f.u.$).[1,2] These anomalies are explained in terms of the 4$f$ band magnetism scenario.[1] The fact that the light rare earth Ce couples antiferromagnetically with the 3$d$ moment is well understood in this scenario. CeFe$_2$ is known to be a ferromagnet (FM) with a fluctuating antiferromagnetic (AFM) ground state. This AFM state gets stabilized by certain substitutions such as Ru, Re, Ir, Al. Ga, Si etc. at the Fe site, thereby giving rise to a FM-AFM transition on cooling.[3-6] It is established that doped CeFe$_2$ compounds can be a model system to understand the physics of metamagnetic transitions, metastability and phase co-existence which are important to understand the properties of certain functional materials which have drawn a lot of interest recently.[6-8] However, a much less attention has been devoted to the substitution at the Ce site by other rare earth elements.[9-13] Enhancement of Curie temperature has been found with $R$ substitution, which enhances the trivalent behavior of Ce in them.[9,12,13] While investigating the effect of various $R$ doping, we have found that Er substitution shows remarkable differences with regard to the magnetic properties, as compared to the other rare earth dopants such as Gd or Ho. These include large magnetic hysteresis at low temperatures and the bifurcation of the zero field cooled (ZFC) and field cooled (FC) magnetization data. So Er is seen to play a totally different role compared to other (heavy) rare earths in doped CeFe$_2$. In order to highlight these special features of Er in CeFe$_2$, we focus on (Ce$_{1-x}$Er$_x$)Fe$_2$ compounds with $x$ = 0.08, 0.12, 0.15, 0.25. We find that Er doping induces the re-entrant spin glass (RSG) state in this series for $x \leq 0.25$.

The physics of spin glasses is still to be resolved and there are a few well known spin glass systems available in reality.[14-16] On the other hand, the reentrant spin glass behavior is realized in a variety of systems such as AuFe, (Eu,Sr)S, FeCr, NiMn, AlFe, (Pd,Fe)$_{1-x}$Mn$_x$, (Eu,Sr)Te, (Eu,Sr)As, amorphous (a-) FeNi, a-FeMn, a-FeCr and a-ZrFe, a-(Fe$_{1-}$ $_x$Mn$_x$)$_{75}$P$_{16}$B$_6$Al, a-(Fe$_{1-x}$Ni$_x$)$_{75}$P$_{16}$BAl$_3$ [17-26]. Recently RSG has been seen in some perovskite manganites,[27-30] shape memory alloys[31,32] and rare earth- transition metal intermetallics[33] as well. RSG behavior occurs when the material shows spin glass behavior at temperatures lower than the ferromagnetic/antiferromagnetic ordering temperature. The observation of RSG behavior is sometimes controversial and ambiguous as similar experimental features appear due to the deviation from perfect ferromagnetic or antiferromagnetic state and/or competition between ferromagnetic and antiferromagnetic phases. So it is necessary to confirm the RSG with proper experimental tools. It is indeed possible to confirm RSG behavior as there are some unique experimental outcomes in the case of RSG.[14,15]

In the present case of Er doped CeFe$_2$ compounds, we discuss the observation of RSG behavior as revealed by dc and ac susceptibility measurements. The distinct features of spin glass behavior below the Curie temperature is established by examining the frequency dependence of susceptibility, relaxation in dc magnetization and remanence present in these compounds.

## II. Experimental Details

Polycrystalline compounds, (Ce$_{1-x}$Er$_x$)Fe$_2$ [$x$ = 0.08, 0.12, 0.15, 0.25] were prepared by arc melting method in a water cooled copper hearth under argon atmosphere. The constituent elements, of at least 99.9% purity, were melted by taking their stoichiometric proportion. The alloys buttons were remelted several times to ensure homogeneity. The arc melted samples were annealed for 10 days in the following way: 600 ºC for 2 days, 700 ºC for 5 days, 800 ºC for 2 days and 850 ºC for 1 day.[5] The structural analysis was performed by the Rietveld refinement of room temperature x-ray diffraction patterns (XRD). The ac magnetic susceptibility data has been carried out in PPMS (Quantum Design) in the frequency range 33-9997 Hz. The ac measurements have also been performed in various ac amplitudes and various dc bias fields. All the ac measurements have been taken during heating after cooling the sample in zero field. The dc magnetization and heat capacity measurements were also performed in the PPMS. DC magnetization has been measured during heating after zero field cooling (ZFC) and field cooling (FC) the sample.

**III. Results**

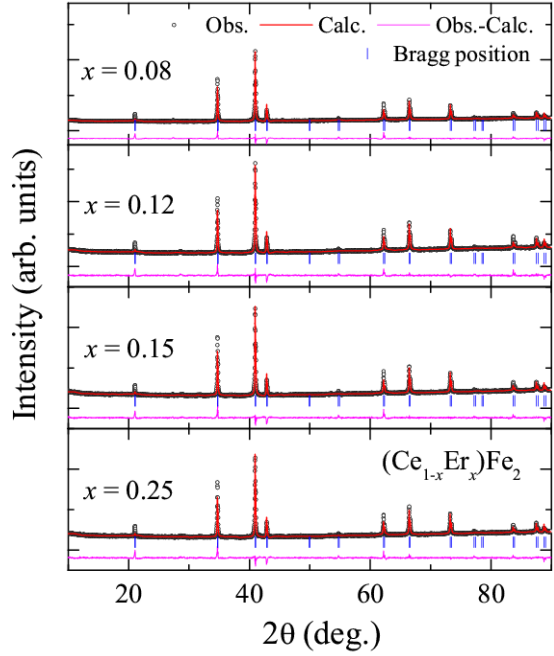

FIG. 1. Room temperature x-ray diffraction patterns of $(Ce_{1-x}Er_x)Fe_2$ [$x$ = 0.08, 0.12, 0.15 and 0.25] compounds, along with the Rietveld refinement.

In Fig. 1, the observed x-ray diffraction patterns along with the calculated patterns and the difference plots are shown. It is clear that the compounds have formed in single phase. At room temperature, these compounds possess the MgCu$_2$ type cubic structure with the space group $Fd\bar{3}m$. The lattice parameter is found to increase from 7.3029(2) Å for $x$ = 0.08 to 7.3047(2) Å for $x$ = 0.25. This variation is in very good agreement with the report by Tang et al.[9]

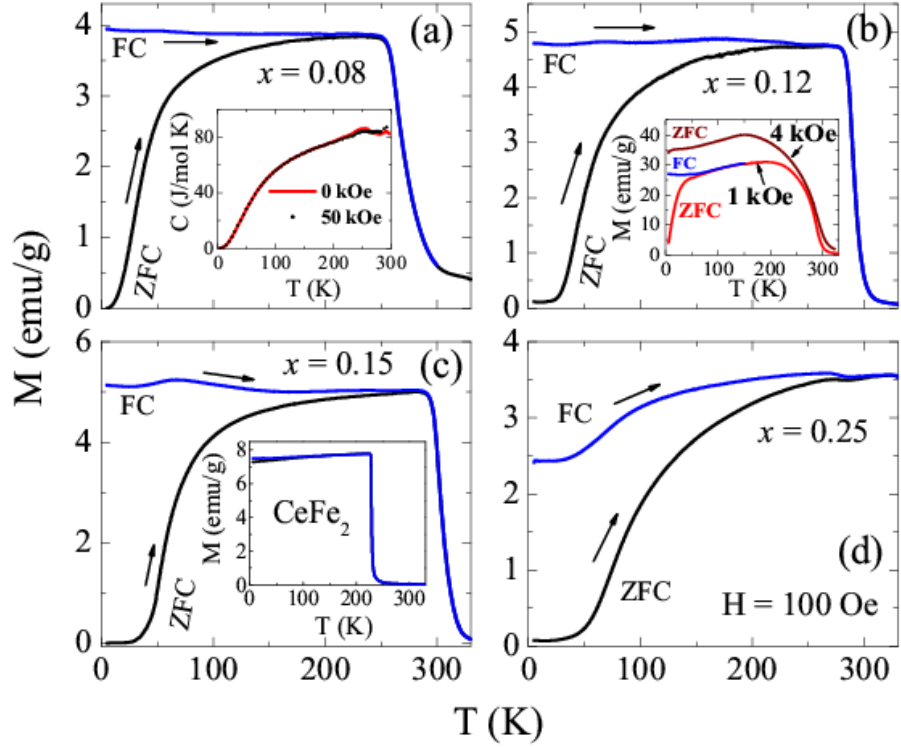

FIG. 2. Temperature variation of dc magnetization of $(Ce_{1-x}Er_x)Fe_2$ [$x$ = 0.08, 0.12, 0.15 and 0.25] compounds during heating in ZFC and FC modes at $H$ = 100 Oe. The inset in (a) shows the variation of heat capacity as a function of temperature (at $H$ = 0 and 50 kOe) in the case of $x$ = 0.08. The inset in (b) shows ZFC and FC magnetization data in 1 kOe and 4 kOe in $x$ = 0.12. The ZFC and FC $M$-$T$ data in the undoped $CeFe_2$ in 100 Oe is shown in the inset of (c).

Figure 2 shows the temperature variation of dc magnetization, $M(T)$, for all the compounds at $H$ = 100 Oe. It can be noticed that all the compounds undergo paramagnetic (PM) to ferromagnetic transition and that the $T_C$ monotonically increases with Er content. Therefore, it is clear that there is a net enhancement of the ferromagnetic ordering with Er. It can be seen that the FC curves reflect the ferromagnetic behavior in all the cases, except in $x$ = 0.25 (see Fig. 2(d)). This latter compound does not follow the FM behavior, possibly due to the fact that the $T_C$ in this case is more than 330 K and that the field cooling is not started from the paramagnetic state. Fig. 2 also shows that at temperatures below $T_C$, the ZFC and FC data follow different paths, resulting in a large bifurcation between them. This behavior roughly indicates the magnetic frustration and glassy behavior at low temperatures ($T < T_C$). It is of interest to note that undoped $CeFe_2$ does not exhibit this bifurcation as can be seen from the inset of Fig. 2(c). In the presence of a higher field (1 kOe), the ZFC-FC difference decreases and at $H = 4$ kOe, the ZFC data shows almost a normal ferromagnetic behavior (see inset of Fig. 2(b)), in the case of $x = 0.12$. Therefore the $M$-$T$ data is indicative of a possible spin glass phase below the Curie temperature in these compounds, which requires further confirmation. As is clear from the inset of Fig. 2(a), the heat capacity does not show any anomaly close to the region where the ZFC magnetization decreases considerably.

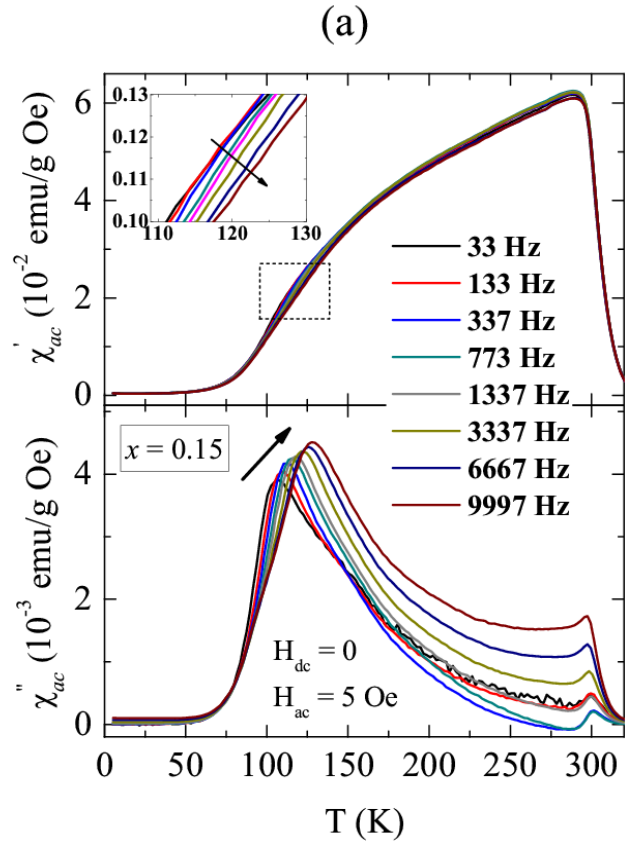

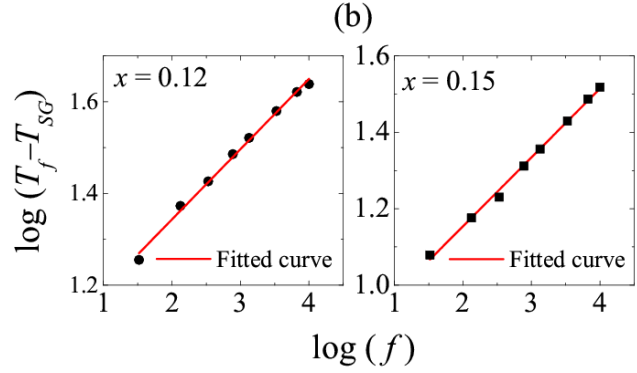

FIG. 3. (a) Temperature variation of in-phase (upper panel) and out-of-phase (lower panel) of ac susceptibility at $H_{ac}$ = 5 Oe for $(Ce_{0.85}Er_{0.15})Fe_2$ at different frequencies. Inset shows the expanded 110-130 K region. (b) Frequency dependence of the freezing temperature in $x$ = 012 and 0.15 along with the fit to the critical slowing down formula, $\tau = \tau_0 (T_f/T_{SG} - 1)^{-zv}$.

To establish the spin glass state, one must study the frequency dependence of the presumed spin glass transition obtained from the dc magnetization measurements. Fig. 3 shows the temperature variation of the in-phase ($\chi'_{ac}$) and the out-of-phase ($\chi''_{ac}$) components of ac magnetic susceptibility for the $x$ = 0.15 compound as a typical example, at different frequencies over a wide temperature range from 5 K to the Curie temperature, at fixed $H_{ac}$ = 5 Oe and $H_{dc}$ = 0. It is to be noted here that when the temperature is reduced below $T_C$, the in-phase component decreases giving a high value at $T_C$ (Fig. 3(a)). On the other hand, the $\chi''_{ac}$ data shows a weak peak at $T_C$, but a pronounced peak is observed close to the temperature at which the ZFC magnetization changes considerably. The decrease of $\chi'_{ac}$ usually indicates the reduction in the ability of the material to respond to the low ac magnetic field. The behavior of $\chi'_{ac}(T)$ is determined by the change of domain wall motion and domain magnetization reorientation in the alternating magnetic field. As the domain magnetization reorientation is insignificant in low (here 5 Oe) ac magnetic field, the magnetization is mainly governed by the domain wall motion. The domain structure originates from the magneto-crystalline anisotropy and is affected by the defects of the crystal lattice.[34] The peak in $\chi''_{ac}$-$T$ plot indicates energy absorption associated with the domain wall motion and domain rotation, which implies that such losses are quite large at temperatures below $T_C$ in the present case.

To further probe the ac response of these compounds, the frequency dependence of the ac magnetic susceptibility was measured, as shown in Fig.3a. It is evident from the inset of Fig. 3a that the $\chi'_{ac}$ data shows strong frequency dependence, with an upward shift of the curve, at around 95-130 K. Similarly, the peak in the $\chi''_{ac}$ data (in the same region) also shows a significant upward shift (lower panel of Fig. 3a). These observations underline the frustrated magnetic state associated with the spin glasses. On the other hand, the frequency dependence of the peak at $T_C$ is found to be quite negligible, for both the in-phase and the out of phase components. These observations strengthen the presumption of a spin glass transition below $T_C$ in this compound. The low frequency (33 Hz) ac susceptibility anomaly coincides with the temperature at which the dc (low field) magnetization falls rapidly. Based on the ac and the dc data, we define this temperature as $T_f$, the freezing temperature. The frequency dependence of $T_f$ is often quantified as $Q = \Delta T_f / [T_f \log_{10}(f)]$,[35,36] which is calculated to be 0.09 and 0.07 for $x = 0.12$ and 0.15 compounds respectively. It is of interest to note that the present values compare well with the values of 0.06 seen in certain shape memory alloys showing RSG state,[31] 0.037 seen in metallic glasses[37] and 0.095 reported in $LaCo_{0.5}Ni_{0.5}O_3$ [38]. On further analysis, the $T_f$ is found to obey the critical slowing down dynamics (see Fig. 3b) governed by the relation $\tau = \tau_0 (T_f / T_{SG} - 1)^{-zv}$, where $\tau$ is relaxation time and $zv$ is known as dynamic exponent.[27] We found the best fit with $T_{SG} = 82$ K, $\tau_0 = 2.7 \times 10^{-7} s$, $zv = 5.55$ for the $x = 0.12$ and $T_{SG} = 95$ K, $\tau_0 = 1.4 \times 10^{-6} s$, $zv = 6.67$ for $x = 0.15$. For a conventional spin glass, $\tau_0$ is $\sim 10^{-10} - 10^{-13}$ s and $zv$ lies in the range of 4-13 [37]. The fact that the present $\tau_0$ values are higher implies that the relaxation is slower and that the RSG phase is constituted by randomly magnetized clusters, instead of atomic level randomness. Such higher $\tau_0$ values have also been found in other RSG systems such as Heusler alloys, $LaCo_{0.5}Ni_{0.5}O_3$, pyrochlore molybdates etc.[31,38,39] It is also found that the magnitude of the peak in $\chi''_{ac}$ increases with increase in frequency. This is again a signature of conventional spin glasses, though some known RSG systems show the opposite trend.[31] It is of importance to mention here that the strong frequency dependence that we have seen in $x = 0.08$, 0.12 and 0.15 is absent in the case of $x = 0.25$.

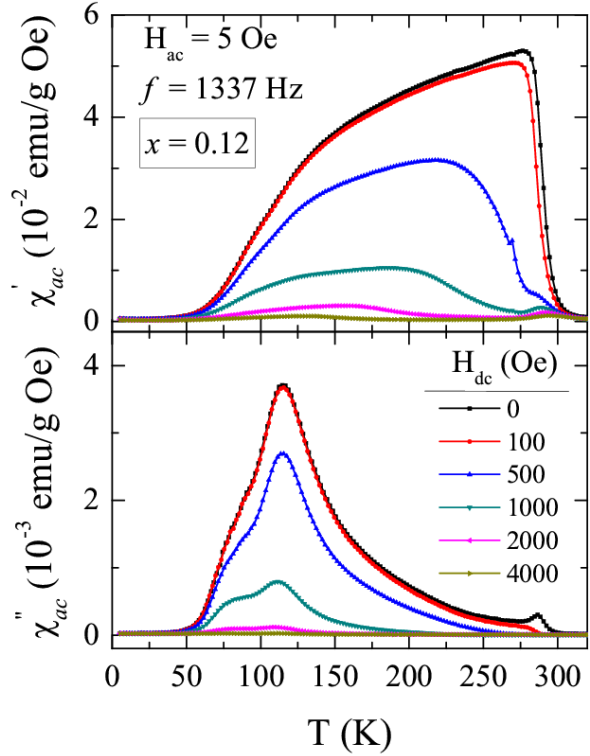

FIG. 4. Temperature variation of (a) in-phase ($\chi'_{ac}$) and (b) out-of-phase ($\chi''_{ac}$) of ac susceptibility for $(Ce_{0.88}Er_{0.12})Fe_2$ at different dc bias fields (0, 100, 500, 1000, 2000, 4000 Oe) at fixed $H_{ac} = 5$ Oe and $f = 1337$ Hz.

By applying a dc bias field ($H_{dc}$), the response of the ac susceptibility diminishes with significant modification in its behavior, which can be seen from Fig. 4. The magnitude of $\chi'_{ac}$ is strongly suppressed with the application of dc bias field. Furthermore, it shows a well defined double hump behavior for $H_{dc} \geq 500$ Oe. The variation of $\chi''_{ac}$ is also more or less similar to that of $\chi'_{ac}$. It is noteworthy that the $\chi''_{ac}$ peak broadens with dc bias field. At $H_{dc} = 4$ kOe, both $\chi'_{ac}$ and $\chi''_{ac}$ become almost zero. This shows that the material is unable to respond at low $H_{ac}$ value (= 5 Oe) when the dc bias field is as large as 4 kOe. Similar findings have been reported in the case of re-entrant spin glass compound $La_{0.96-y}Nd_yK_{0.04}MnO_3$ [28].

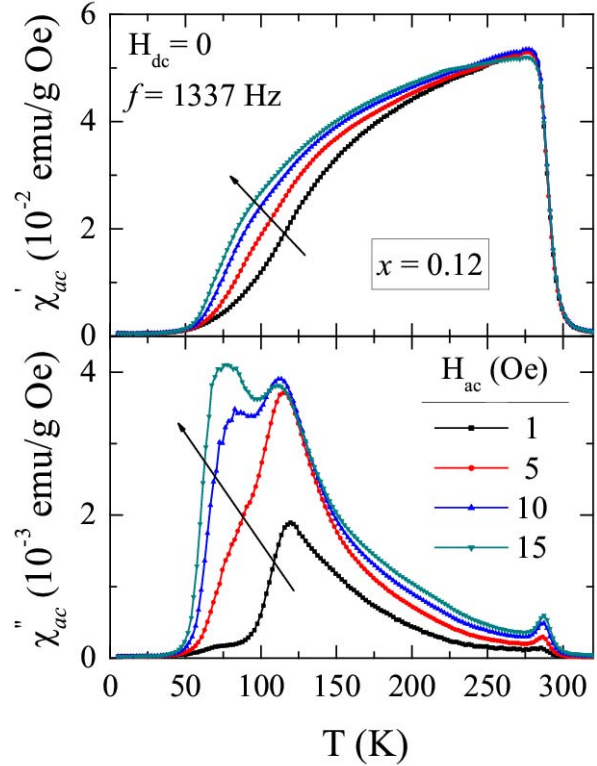

FIG. 5. Temperature variation of (a) in-phase ($\chi'_{ac}$) and (b) out-of-phase ($\chi''_{ac}$) of ac susceptibility for $(Ce_{0.88}Er_{0.12})Fe_2$ at different ac field amplitudes (1, 5, 10 and 15 Oe) at a fixed $f$ =1337 Hz and $H_{dc}$ = 0.

The effect of amplitude of the alternating field on the ac susceptibility has been illustrated in Fig. 5. As expected, a huge enhancement and modification of the both in-phase and out-of-phase susceptibility can be seen near the freezing temperature. It can also be noted that the changes around the $T_C$ region are quite nominal. With increase in the ac field, the important observations are (i) both in-phase and out-of-phase peaks shift towards lower temperatures and (ii) peak in the out-of-phase part broadens and splits into two. The first observation is consistent with the fact that higher ac magnetic amplitude weakens the occurrence of spin glass state which in turn shifts the $T_f$ towards lower temperature. The second observation can be attributed to the complex magnetic state in the RSG phase. It is possible that the double peak, which develops at higher dc (Fig. 4) or ac (Fig. 5) fields is due to the additional contribution from a few isolated moments which are not part of any cluster.

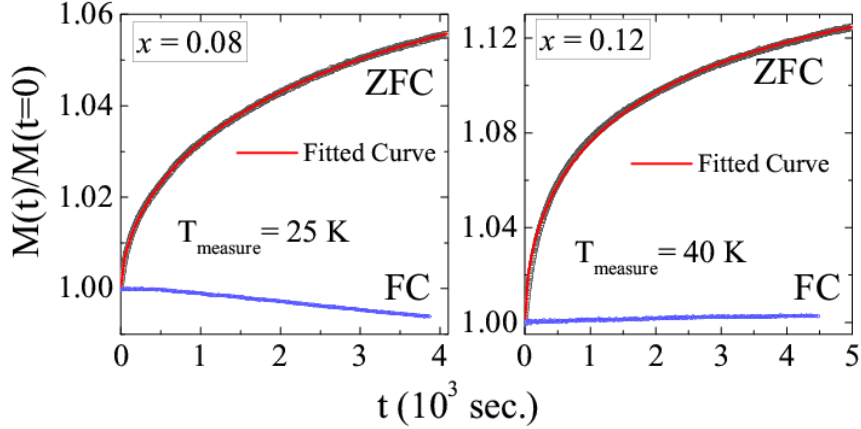

FIG. 6. Time relaxation of magnetization, $M(t)$ for $(Ce_{0.92}Er_{0.08})Fe_2$ and $(Ce_{0.88}Er_{0.12})Fe_2$ at $(T_{measure}, H_{dc})$ = (25 K, 500 Oe) and (40 K, 100 Oe), respectively. Data has been taken during ZFC and FC modes. Red line shows the typical relaxation fit of the type: $M_t(H) = M_0(H) + [M_\infty(H) - M_0(H)][1 - \exp\{-(t/\tau)^\alpha\}]$.

All the features presented above clearly point towards glassiness at low temperatures. In order to further ascertain whether the magnetic state is indeed of spin glass type, the magnetization relaxation measurement has been carried out in detail. Comparison of the time dependent ZFC and FC magnetization data can be used as a tool to investigate the evolution of the spin configuration and possible occurrence of re-entrant spin glass (RSG) phase.[28] For this, the sample is cooled in presence (FC) or absence (ZFC) of a field to the measurement temperature. Then the magnetic field was applied (in ZFC case) and then the time variation of the growth of the magnetization is recorded. The ZFC magnetization measured in this way shows a huge relaxation at $T = 25$ K (for $x = 0.08$) and 40 K (for $x = 0.12$) (as shown in Fig. 6), which indicates the metastability of the low temperature magnetic state. However, there is no considerable relaxation observed in the FC magnetization. This is in sharp contradiction with a magnetic glass where FC magnetization relaxes and ZFC shows no relaxation.[40] This difference seen in Fig. 6 clearly shows that these compounds are not magnetic glasses like $Ce(Fe_{0.96}Ru_{0.04})_2$, but re-entrant spin glasses. In the present compounds, the ZFC magnetization at a constant field and temperature grows as a function of time and the growth can be fitted well to a stretched exponential of the type: $M_t(H) = M_0(H) + [M_\infty(H) - M_0(H)][1 - \exp\{-(t/\tau)^\alpha\}]$, where $\tau$ is the characteristic relaxation time and $\alpha$ is called stretching parameter that ranges between 0 and 1. Best fit curve gives $\alpha = 0.58$ and $0.53$ for the $x = 0.08$ and $0.12$ compounds respectively.

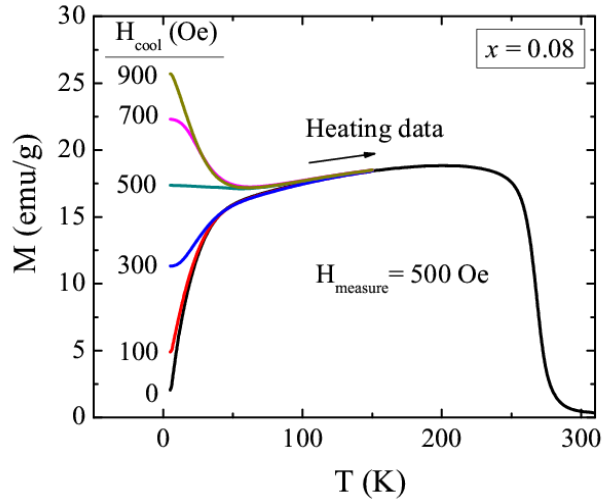

FIG. 7. Temperature dependence of magnetization for $(Ce_{0.92}Er_{0.08})Fe_2$ during heating in a field of 500 Oe after cooling the sample in different magnetic fields of 0, 100, 300, 500, 700 and 900 Oe.

Yet another evidence in favor of RSG state in the present case is the result obtained from the dc measurement technique proposed by Roy et al.[40] This protocol involves the measurement of $M(T)$ after cooling and heating the sample in unequal fields (see Fig. 7). We have chosen the compound with $x = 0.08$ as the prototype sample for this study. When the cooling field is more than the measuring field ($H_{cool} > H_{measure}$), the magnetization starts decreasing from a higher value as the temperature increases. On the other hand, when the cooling field is less than the measuring field ($H_{cool} < H_{measure}$), the trend reverses at low temperatures. The observation in the present case is similar to that in reentrant spin glass systems such as $Au_{82}Fe_{18}$, but in sharp contrast to the magnetic glass namely $Ce(Fe_{0.96}Ru_{0.04})_2$ [40]. This clearly shows that the glassiness observed with Er substitution is quite different from that seen in Fe site doped $CeFe_2$.

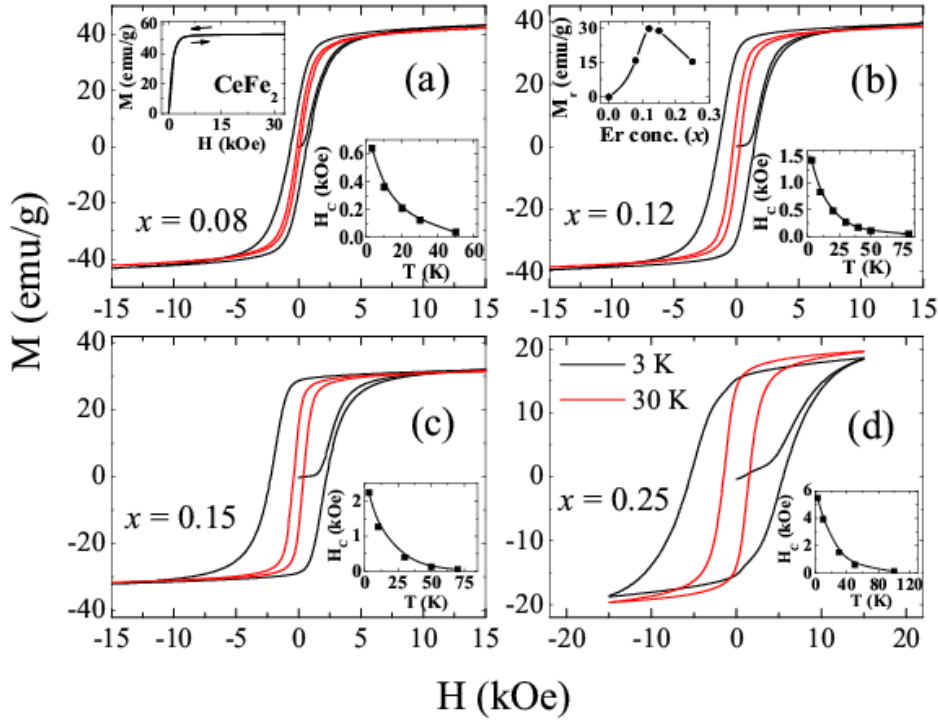

FIG. 8. *M-H* plots of $(Ce_{1-x}Er_x)Fe_2$ [$x$ = 0.08, 0.12, 0.15 and 0.25] compounds at 3 K and 30 K. The insets at lower right corners show the variation of coercive fields ($H_C$) with temperature. The inset at upper left corner of (a) shows the two loop *M(H)* plot for $CeFe_2$ at 3 K. The inset at upper left corner of (b) shows the variation of remanent magnetization ($M_r$) at 3 K with Er concentration.

Based on the observations presented above, it is quite evident that the Er doped compounds enter a spin glass state as they are cooled from $T_C$. Another important characteristic of this RSG state is the presence of considerable remanence and coercivity.[27] To probe this, the *M-H* plots have been recorded in all the compounds both under ZFC and FC modes. It is found that there is no difference in the data between these two modes. Magnetization isotherms, which are presented in Fig. 8, show clear hysteresis at $T$ = 3 K and 30 K which was not observed in undoped $CeFe_2$ (see inset at the upper corner of Fig. 8(a)). We would like to highlight here that no such hysteresis was observed with other rare earths such Gd or Ho even at the lowest temperature. Tang et al. have shown that even with Er, the hysteresis is visible only when the Er concentration is below a critical value ($x = 0.7$).[9] Though Gd substitution does not alter the magnetocrystalline anisotropy considerably, replacing Ce with Ho should have increased the net rare earth sublattice anisotropy. The absence of hysteresis in both Gd and Ho compounds positively indicates that the hysteresis seen in the present (Er doping) case is not due to the increase in the magnetocrystalline anisotropy usually expected with Er addition. We would like to emphasize that even in the field cooled mode, the *M-H* loops are all symmetric with respect to both the *M* and *H* axes, thereby indicating the absence of any exchange bias/exchange anisotropy.

The magnetization is found to be saturated in all the compounds at around 15 kOe and the saturation value is almost unchanged between 3 and 30 K, except in $x = 0.25$. In this compound ($x = 0.25$), the saturation magnetization is found to be more at 30 K as compared to that at 3 K, indicating the predominant ferrimagnetic coupling in this compound. It may be noted that the *M-T* curve in the FC mode (Fig. 2) also indicated this behavior, though it is true that the field cooling did not start from the paramagnetic phase. But since the *M-H* plots in the other Er concentrations showed no difference between the FC ad ZFC modes, we can take the anomalous *M-H* behavior of $x = 0.25$ as real. At this point, it may be recalled that the frequency dependence of ac susceptibility was almost negligible in this compound. With increase in temperature, the hysteresis is found to decrease as can be seen from the decrease in the coercive field ($H_C$) shown in the insets at lower right corners of Fig. 8. The value of $H_C$ at 3 K for $x = 0.08$ and 0.25 are found to be 641 Oe and 5.5 kOe, respectively. The remanent magnetization ($M_r$) is found to increase initially with Er concentration and then decrease at $x = 0.25$ (see the inset at the upper left corner of Fig. 8(b)). We have also observed that at $x = 0.5$, the *M-H* curve at 3 K shows a metamagnetic transition (not shown), similar to that seen in other reports.[9] It has also been reported that $(Ce_{1-x}Tb_x)Fe_2$ as well as $(Ce_{1-x}Dy_x)Fe_2$ shows metamagnetic transition around $x = 0.5$ [12,13].

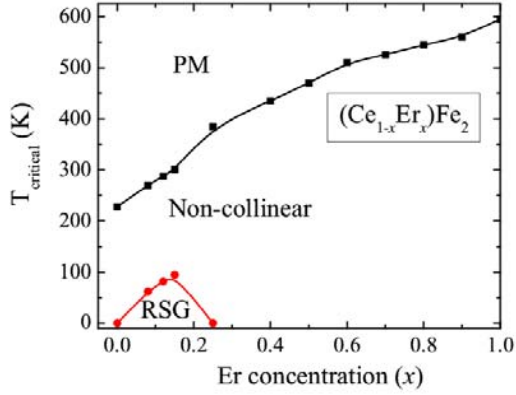

FIG. 9. The magnetic phase diagram of $(Ce_{1-x}Er_x)Fe_2$ system. $T_{critical}$ refers to the Curie temperature or the spin glass freezing temperature.

Taking into account the temperature variation of the magnetic state as revealed by the results presented above, we have proposed the magnetic phase diagram for $(Ce_{1-x}Er_x)Fe_2$, as shown in Fig. 9. The $T_C$ values for $x \geq 0.25$ compounds have been taken from the ref 9. The phase diagram shows that within a range of Er concentration, RSG state appears. In the region of $T_{SG} \leq T \leq T_C$, the system seems to deviate from a perfect ferrimagnet and we define it as a non-collinear ferrimagnetic state, and ultimately, $ErFe_2$ shows a normal ferrimagnetic behavior.

## IV. Discussion

Occurrence of RSG state in $(Ce_{1-x}Er_x)Fe_2$ is rather unexpected. The RSG state is found to be a result of randomly magnetized clusters, instead of random atomic moments. The fact that it is not achieved with rare earths such as Ho or Gd makes it even more interesting to probe its origin. An experimental observation that has been seen with various rare earth dopings in $CeFe_2$ is that the $T_C$ increases with the rare earth content. This implies that there is an increase in the net ferromagnetic coupling with the addition of $R$, even when the low temperature state shows spin glass signatures. It should also be noted that the moment on Ce in undoped $CeFe_2$ is quite small, due to the delocalized character of the 4$f$ shell. In fact this is the main reason for the low $T_C$ of $CeFe_2$. Substitution of rare earths at Ce site is found to change the valence state of Ce.[9,12,13] It has been reported that Ce valency in CeFe$_2$ is close to 3.29, which decreases to almost 3 on $R$ substitution, resulting in an increase in the Ce moment. Because of the reduction in the delocalization of Ce, the Fe sublattice moment would also increase, resulting in the increase in $T_C$ because of the increase in the $3d$- $3d$ exchange. Since Er occupies random (Ce) positions in the unit cell, it is quite likely that the Ce moment is non-uniform in the Er doped compounds. It is also known that the coupling of both Ce and heavy rare earths, with Fe is anti parallel. All these point towards the fact that the magnetic structure in doped compounds, especially with smaller $x$ values, is non-collinear. One can attribute the magnetic glassiness and the frustration to this non-collinearity. But the interesting point is that though Er doping resulted in the RSG state, the Ho doping did not show any evidence of it, though the $T_C$ and the lattice parameter variations are identical to those of Er. In this context, it is to be noted that ErFe$_2$ is the only member in the RFe$_2$ series which shows magnetic compensation (at 468 K) in the $M$-$T$ data. The (Ce$_{1-x}$Er$_x$)Fe$_2$ shows compensation below 300 K.[9] However, there are no reports of such compensation in Gd or Ho doping. Therefore, it seems that the formation of spin glass state depends on whether the Fe and ($R$+Ce) subalttice moments are comparable or not. If they are comparable, it is reasonable to assume that the direct FM coupling of the Fe sublattice competes with the indirect AFM coupling between Fe and ($R$+Ce) moments. The fact that above a certain concentration of Er, the compound behaves more or less like a normal ferrimagnet supports this proposition. Another noteworthy point is the absence of any exchange bias in the FC magnetization isotherms. This probably indicates that in the RSG state, there is no FM component, unlike in some materials in which the spin glass state is assumed to coexist with the FM state.[31] It should be mentioned here that, like Er substitution, certain concentration of Tb and Dy doped in CeFe$_2$ [namely Ce$_{0.6}$Tb$_{0.4}$Fe$_2$ and Ce$_{0.6}$Dy$_{0.4}$Fe$_2$, respectively] is reported to show a compensation point, with the former showing the compensation below 300 K.[13]

Finally, we would like to compare the findings of this work with those of Er$_{0.75}$Dy$_{0.25}$Al$_2$, which was reported recently.[34] The authors have reported many observations similar to those of the present system and attributed them to the glassiness at low temperatures. They have suggested that the competing anisotropies of the randomly distributed Er and Dy ions are responsible for the glassiness. However, we would like to emphasize here that the frequency dependence of the ac magnetic susceptibility in the present case is much more pronounced than in $Er_{0.75}Dy_{0.25}Al_2$. Furthermore, unlike the present case, the coercivity at 2 K was almost negligible in $Er_{0.75}Dy_{0.25}Al_2$. Therefore, we feel that the scenario in our case is different and indeed that of a re-entrant spin glass. However, a detailed study with some local probes is essential to confirm this presumption.

**V. Conclusions**

In this report we have shown that $(Ce_{1-x}Er_x)Fe_2$ [$x$ = 0.08, 0.12, 0.15, 0.25] compounds show re-entrant spin glass behavior. Frequency dependence of ac susceptibility, relaxation in ZFC dc magnetization below the freezing temperature and intrinsic remanence are shown to establish the re-entrant spin glass state. We show here that the freezing temperature follows the critical slowing down mechanism and that the parameters obtained are comparable to those of typical spin glasses and other known RSG systems. We further conclude that the occurrence of RSG state in this system is dependent on Er concentration. The RSG state is attributed to the random distribution of Er and the modification of the Ce and Fe moments due to the band structure changes brought about by the lattice expansion. The RSG state exists over certain concentrations of Er. The RSG state is found to be associated with the randomly magnetized clusters, instead of atomic level randomness. With increase in Er concentration, the system is found to gradually change to a non-collinear ferrimagnet and finally to a normal ferrimagnet at $x$ = 1. The fact that all rare earths do not give rise to these features in $CeFe_2$ makes this study interesting.

*Acknowledgement*

KGS and AKN thank Board of Research on Nuclear Sciences (Dept. of Atomic Energy, Govt. of India) for providing the financial support for carrying out this work.

**References**

[1]O. Eriksson, L. Nordstrom, M. S. S. Brooks, and B. Johansson, Phys. Rev. Lett. **60**, 2523 (1988).

[2]C. Giorgetti, S. Pizzini, E. Dartyge, A. Fontaine, F. Baudelet, C. Brouder, P. Bauer, G. Krill, S. Miraglia, D. Fruchart and J. P Kappler, Phys. Rev. B **48,** 12732 (1993).

[3]S. B. Roy and B. R. Coles, Phys. Rev. B **39**, 9360 (1989).

[4]K. Koyamaa, K. Fukushimaa, M. Yamadab, T. Gotob, Y. Makiharac, H. Fujiid and K. Watanabe, Physica B **346-347**, 187 (2004).

[5]A. Haldar, K. G. Suresh and A. K. Nigam, Phys. Rev. B **78**, 144429 (2008).

[6]A. Haldar, N. K. Singh, Ya. Mudryk, K.G. Suresh, A.K. Nigam and V.K. Pecharsky, Solid State Communications **150**, 879-883 (2010).

[7]S. B. Roy, G. K. Perkins, M. K. Chattopadhyay, A. K. Nigam, K. J. S. Sokhey, P. Chaddah, A. D. Caplin, and L. F. Cohen, Phys. Rev. Lett. **92**, 147203 (2004).

[8]S. B. Roy, P. Chaddah, V. K. Pecharsky, Jr. K. A. Gschneidner., Acta Materialia **56**, 5895 (2008).

[9]C. C. Tang, Y. X. Li, J. Du, G. H. Wu and W. S. Zhan, J. Phys.: Condens. Matter **11**, 2027 (1999).

[10]S. F. Cunha, A. P. Guimaraes and F. P. Livi, J. Phys. Chem. Solids **41**, 761 (1980).

[11]E. T. Miskinis, K. S. V. Narsimhan, W. E. Wallace and R. S. Craig, J. Solid State Chem. **13**, 311 (1975).

[12]C. C. Tang, D. F. Chen, Y. X. Li, G. H. Wu, K. C. Jia and W. S. Zhan, J. Appl. Phys. **82**, 4424 (1997).

[13]C. C. Tang, W. S. Zhan, D. F. Chen, Y. X. Li, J. Du, B. G. Shen and G. H. Wu, J. Phys.: Condens. Matter **10**, 2797 (1998).

[14]J. A. Mydosh, J. of Magn. and Magn. Mat. **7**, 237 (1978).

[15]J. A. Mydosh, Spin Glasses: An Experimental Introduction (Taylor & Francis, London, 1993).

[16]K. Binder and A. P. Young, Rev. Mod. Phys. **58**, 801 (1986).

[17]G. Aeppli, S. M. Shapiro, R. J. Birgeneau, H. S. Chen, Phys. Rev. B **29**, 2589 (1984).

[18]M. Hennion, B. Hennion, I. Mirebeau, S. Lequien and F. Hippert, J. Appl. Phys. **63**, 4071 (1988).

[19]S. Lequien, B. Hennion, and S. M. Shapiro, Phys. Rev. B **38**, 2669–2674 (1988).

[20]S. K. Burke, R. Cywinski, and B. D. Rainford, J. Appl. Crystallogr. **11**, 644 (1978).

[21]K. Motoya, S. M. Shapiro and Y. Muraoka, Phys. Rev. B **28**, 6183 (1983).

[22]Wei Bao, S. Raymond, S. M. Shapiro, K. Motoya, B. Fak, and R. W. Erwin, Phys. Rev. Lett. **82**, 4711–4714 (1999).

[23]B. H. Verbeek, G. J. Nieuwenhuys, H. Stocker, and J. A. Mydosh, Phys. Rev. Lett. **40**, 586 (1978).

[24]M. B. Salamon, K. V. Rao, and H. S. Chen, Phys. Rev. Lett. **44**, 596 (1980).

[25]S. M. Bhagt, J. A. Geohegan, and H. S. Chen, Solid State Commun. **36**, 1 (1980).

[26]H. Maletta and P. Convert, Phys. Rev. Lett. **42**, 108 (1979).

[27]J. Dho, W. S. Kim, and N. H. Hur, Phys. Rev. Lett. **89**, 027202 (2002).

[28]R. Mathieu, P. Svedlindh, and P. Nordblad, Europhys. Lett. **52**, 441 (2000).

[29]S. H. Chun, Y. Lyanda-Geller, M. B. Salamon, R. Suryanarayanan, G. Dhalenne, and A. Revcolevschi, J. Appl. Phys. **90**, 6307 (2001).

[30]D. N. H. Nam, R. Mathieu and P. Nordblad, N. V. Khiem and N. X. Phuc, Phys. Rev. B **62**, 8989 (2000).

[31]S. Chatterjee, S. Giri, S. K. De, and S. Majumdar, Phys. Rev. B **79**, 092410 (2009).

[32]Eduard Obrado, Antoni Planes, Benjamin Martinez, Phys. Rev. B **59**, 11450 (1999).

[33]Fang Wang, Jian Zhang, Yuan-fu Chen, Guang-jun Wang, Ji-rong Sun, Shao-ying Zhang, and Bao-gen Shen, Phys. Rev. B **69**, 094424 (2004).

[34]E. M. Levin, V. K. Pecharsky and K. A. Gschneidner Jr., J. of Appl. Phys. **90**, 6255 (2001).

[35]R. Mahendiran, Y. Breard, M. Hervieu, B. Raveau and P. Schiffer, Phys. Rev. B **68**, 104402 (2003).

[36]M. K. Singh, W. Prellier, M. P. Singh, R. S. Katiyar, and J. F. Scott, Phys. Rev. B **77**, 144403 (2008).

[37]Q. Luo, D. Q. Zhao, M. X. Pan and W. H. Wang, Appl. Phys. Lett. **92**, 011923 (2008).

[38]M. Viswanathan and P. S. Anil Kumar, Phys. Rev. B **80**, 012410 (2009).

[39]N. Hanasaki, K. Watanabe, T. Ohtsuka, I. Kezsmarki, S. Iguchi, S. Miyasaka, and Y. Tokura, Phys. Rev. Lett. **99**, 086401 (2007).

[40]S. B. Roy and M. K. Chattopadhyay, Phys. Rev. B **79**, 052407 (2009).